\documentclass[aps,pre,twocolumn,showpacs,howkeys,preprintnumbers,amsmath,amssymb]{revtex4}
\usepackage{graphicx}
\usepackage[hang,nooneline]{subfigure}
\usepackage{amsmath}
\usepackage[nooneline]{caption}
\usepackage[latin1]{inputenc}
\usepackage[english,brazil]{babel}
\usepackage{paralist}


\begin{document}

\title{Magnetic Properties of the Metamagnet Ising Model in a three-dimensional Lattice in a Random and Uniform Field}
\author{J. B. dos Santos-Filho}%
\thanks{J. B. Santos-Filho is grateful to CAPES for partial support.}
\author{Douglas F. de Albuquerque}
\email{douglas@ufs.br, douglas.dieb@gmail.com} 
\author{N. O. Moreno}
\thanks{N. O. Moreno acknowledges the support from CNPq and FAPITEC.}
%
\affiliation{Departamento de F\'{i}sica, Universidade Federal de
Mato Grosso, 78060-900, Cuiab\'{a}, MT, Brazil.
}

\date{\today}

\begin{abstract}
By employing the Monte Carlo technique we study the behavior of
Metamagnet Ising Model in a random field. The
phase diagram is obtained by using the algorithm of Glaubr in a cubic lattice of linear size $L$ with values ranging from $16$ to $42$ and with periodic boundary conditions.%
\end{abstract}

\maketitle


\section{Introduction}

%
Random fields and disordered magnetic systems have been a considerable source of research in recent years.\cite{Belanger2000} The random field Ising model (RFIM) has been one of the most interesting subject of research in Physics of Condensed Matter in the last fifteen years and it occupies prominence position when we deal with disordered systems.\cite{Belanger1998, Birgeneau1998} The types most common of disorder are represented by i) disorder in the bonds and ii) randomness in the strength of the applied magnetic field. In this model the disorder depends on the applied external magnetic field and although RFIM has deserved many investigations from both experimental and theoretical points of view,\cite{wc1992S} no conclusive result has been achieved for the understanding the nature of the phase transitions and critical behavior. On the other hand, questions as the lower critical dimension~\cite{Imbrie1984, prl351975I} and the existence of a static phase transition have already been solved from the theoretical point of view. However, questions as the existence of the tricritical point are still opened.\cite{prb391989A} The relevance of RFIM is due to the fact that it is the simplest to describe the essential physics of various class of disordered systems, which includes: \begin{inparaenum}[(i)]\item structural phase transitions in random alloys,\cite{Childress1991} \item commensurate charge-density-wave systems with impurity pinning,\cite{Fisher1983, Fisher1985} \item binary fluid mixtures in random porous media,~\cite{Maher1984} \item melting of intercalates in layered compounds such as TiS2,\cite{Suter1982} \item frustration introduced by the disorder in interacting many body systems, besides explaining several aspects of electronic transport in disordered insulators\cite{Efros1975} and \item systems near the metal-insulator transition.\cite{Kirkpatrick1994, Pastor1999}\end{inparaenum} In the last years, the physics of the hysteresis, of the avalanche behavior and of the origin of self-organized criticality\cite{Pastor1999} have been modelled employing the non-equilibrium behavior of the RFIM. In particular, a new class of problems, such as self-generated glassy behabior, has been studied through the non-disordered model with infinitesimal random field.~\cite{Mezard1999} Recently, random magnetic fields have been applied also in metamagnet systems like Ising model and importants results have been obtained.~\cite{Weizenmann2007, Santos2007} An ideal metamagnet crystal can be constructed overlapping identical layers of spin with ferromagnetic coupling between spins next neighbors of each layer and with antiferromagnetic coupling next neighbors spins of the adjacent layers. One another way to think a metamagnet crystal is to consider a cubic crystal with the couplings between the first neighbors as being of antiferromagnetic character and the couplings between the second neighbors as being ferromagnetic. In these systems, only the competition between the ferromagnetic and antiferromagnetic ordering are interesting. However, the application of a random and uniform magnetic field can yield the appearance of new phenomena and a richer critical behabior becomes possible and, in particular, in the present work we study the RFIM applied to a metamagnet system in a cubic lattice by employing Monte Carlo Method.
\section{Model and Simulation}
Metamagnet ideal can be considered as a set of spins with uniaxial anisotropy ferromagnetic interactions within
each  ($J1 > 0$) and antiferromagnetic
interaction $ (J_2<0)\,.$ The hamiltonian model for spin-$1/2$ is given by:
\begin{equation}
\mathcal{H}=-\sum_{\langle i,j\rangle}J_1\sigma_i\sigma_j 
            -\sum_{\langle i,j\rangle}J_2\sigma_i\sigma_k 
            -\sum_{i=1}^N (h-h_i)\sigma_i  \label{hamil}\,,
\end{equation}
where the first is executed on all pairs of spin nearest-neighbors on same plane and second sum run over all pairs of spin nearest-neighbor in parallel planes, $h$ is the strength of the external uniform magnetic field and $h_i$ is the random magnetic field which obeys the bimodal distribution given by:%
\begin{equation}
P(h_i)= \frac{1}{2}[\delta (h_i-h_r) + \delta (h_i + h_r)]\,, 
\end{equation}
where $h_r$ is the strength of the random field.

To study this system we employed Monte Carlo simulation technique~\cite{Landau2000} by using the algorithm of Glauber in a cubic lattice of linear size $L$ with values ranging from $16$ to $42$ and with periodic boundary conditions. To reach the equilibrium state we take, for guarantee, at least $2 \times 10^4$ Monte Carlo steps (MCs) for all the lattice sites we studied and more $3\times 10^4$ MCS to estimate the average values of the quantities of interest. In our work we consider one MCs equivalent $L^3$ trials for change the state of a spin of the lattice.%
 
We calculated the sublattice magnetization per spin belong to the different planes by using

\begin{equation}
m_A= \left[\frac{2}{N}\left\langle \sum_{i\, \in \, A}\sigma_i \right\rangle\right]\,, 
\end{equation}

\begin{equation}
m_B= \left[\frac{2}{M}\left\langle \sum_{i\, \in \, B}\sigma_i \right\rangle\right]\,,
\end{equation}%
The transition lines of the phase diagram were obtained from the staggered magnetization, $m_s$ and magnetization $m$. Thus we calculate $m_s =m_A-m_B$ and $ m= m_A+m_B$ that are our parameters of order for antiferromagnetic and ferromagnetic phase, respectively. In the above equations $\left[\,\cdots\,\right]$ denotes the average over the disorder and $\left< \,\cdots\,\right> $ denotes the thermal average.%
%
\section{Results and Conclusions}

The Figure 1 illustrate the complete phase diagram, for two selected values of the random field $h_r$,  showing the continuous and discontinuous transition lines separating the antiferromagnetic and paramagnetic phases. The tricritical point, which is indicated by an open square, joins these two lines. The points of the phase diagram are obtainded from the knowledge of the point of maximum of the curve for the susceptibility, see Figure 2.It was obtained of the  results by fixing H and changing T. In the case of the discontinuous transition, we determined the magnetization curve as a function of the field, for a fixed value of temperature Figure . This procedure is not an efficient to localize the tricritical point  one, because it is difficult to distinguish a continuous from a discontinuous curve, especially near the tricritical point, but it gives an idea of the range of values of the field where the transition is of first order. In this work the location of the tricritical point was achieved through the disappearance of hysteresis ~\cite{Landau1976}. For a fixed value of temperature, we drew the magnetization curves for increasing and decreasing values of the magnetic field. 

(In Figs., we show these curves for a system of size $L=40$, and for three values of temperature near the tricritical point. Our estimate for the tricritical temperature is T=.)

On the other hand, Figure 3 shows the behavior of the magnetization $m$ and the staggered magnetization $m_s$ as function of the temperature for different values of the field external uniform.%
In this diagram we can observe as the random field affects the behavior of the system.
\begin{figure}[h]
 \includegraphics[scale=0.8]{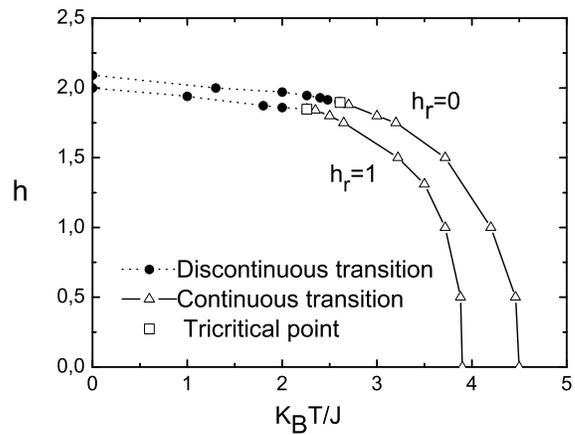}
\caption{Phase diagram of metamagnet in a cubic lattice in the plane $t - h$ for $h_r =0$ and $h_r=1$.}
\end{figure}

\begin{figure}[h]
 \includegraphics[scale=0.77]{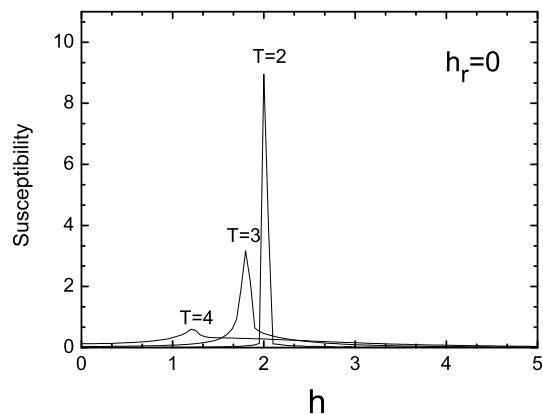}
 \caption{Susceptibility curve {\it versus} temperature, $K_BT/J$, for different values of external field.}
\end{figure}
\begin{figure}[h]
\includegraphics[scale=0.8]{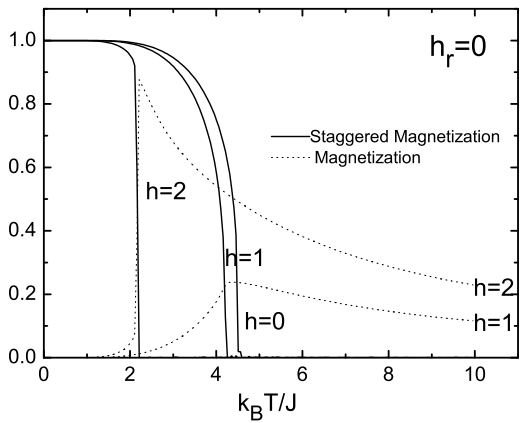}
\caption{Curves of magnetization $m$ and staggered magnetization $m_s$ for different values of external field.}
\end{figure}

\begin{figure}[h]
\includegraphics[scale=0.8]{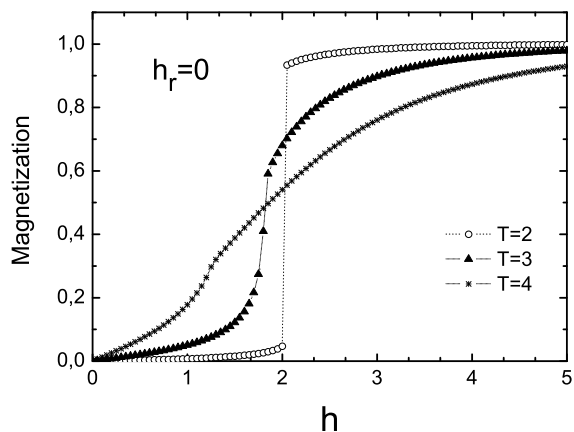}
\caption{Susceptibility curves {\it versus} external field $h$ for different temperatures.}
\end{figure}
%
%
%
%
The behavior of the magnetization when the external field varies for different values of the temperature is shown in Figure 4. For $T=2$ one get a first-order transistion and for $T =3$ and $T =4$ a transistion of second order.

We can observe in Figure $5$ the variation of the susceptibility with the external field for different temperatures. In Figure $6$ we can see better as the random field affects the order of the system, in this Figure we kept the temperature and we calculate the magnetization with the variation of the field uniform for different values of random field. 
\begin{figure}[h]
\includegraphics[scale=0.8]{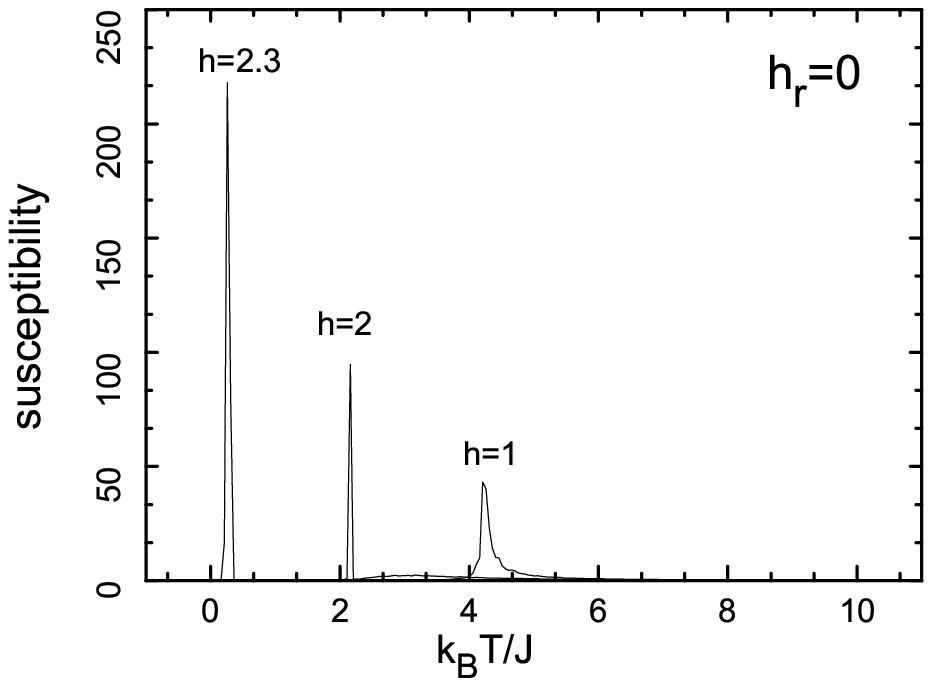}
 \caption{Curves of magnetization $m$ versus external field for different values of temperature.}
\end{figure}

\begin{figure}[h]
\includegraphics[scale=0.8]{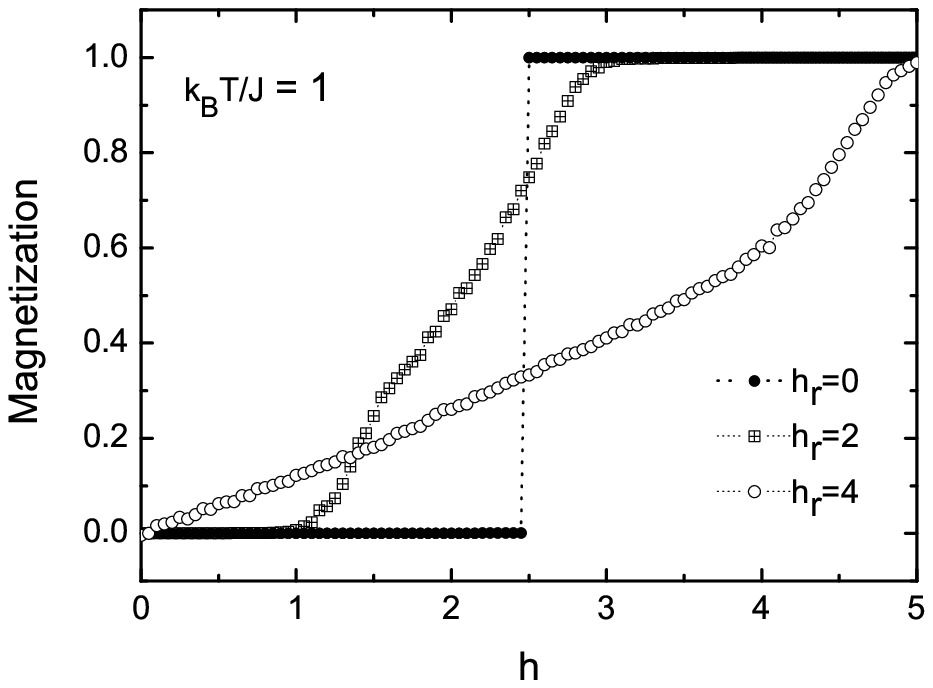}
\caption{Curves of magnetization $m$ versus external field for different values of random field $h_r$.}
\end{figure}
%


In summary, the present Monte Carlo simulations for a metamegnet Ising model in a random and uniform field show that the phase diagram in the plane uniforme field $h$ {\it versus} temperature present continuous and first-order transition lines separated by possible tricritical points.%
%
%
%

\end{document}